\documentclass{article}
\usepackage{amsmath}
\newcommand{\ket}[1]{| #1 \rangle}

\newcommand{\msl}[1]{\mbox{\sl #1}}
\title{The state $\ket{IV}$ (Babel of entanglement)}
\author{\em Alexander Yu.\ Vlasov}
\begin{document}
\maketitle
\begin{abstract}
 Maybe active discussions about entanglement in quantum 
information science demonstrate some immaturity of 
this rather young area. 
So recent tries to look for more accurate ways of 
classification devote rather encouragement than criticism.
\end{abstract}

%\section{Discussion}

Some tries of deeper analyze of conception of entanglement recently 
cased an active criticism \cite{vE}. Really, the examples with quantum
optics\footnote{Photon --- is massless relativistic particle with spin 1.} 
and the vacuum entanglement maybe too complicated for such questions.
Oversimplified treatment of a behaviour of a quantum systems may be 
relevant with even simpler examples. 

A problem --- are not very accurate constructions used in many works 
on quantum information science in relation with the definition of a compound 
quantum system. It may be discussed from both mathematical and physical 
point of view.

From mathematical point of view, in about 99\% of quantum computing papers 
I saw, is given ``simplified'' definition of the tensor product\footnote{
The other fresh ``invention'' is tensor product of {\em Lie groups}, 
like ``$SU(2)\otimes SU(2)$''!}. 
An accurate definition of the tensor product uses a quotient of some 
infinite-dimension space \cite{SLang}. Such a definition could be considered 
as too cumbersome for using in the quantum computing, but it defines tensor
product in {\em basis-independent way}.

Such a basis independent view is important also in the physical applications.
Let us consider two spin-half systems. There is a quite relevant
note of Feynman \cite[\textbf{12-1}]{FLP3} in relation with
hydrogen atom, {\em i.e.}, the system with the proton and electron.
\begin{quotation}
\small
 The first question we have to answer is: What are the {\em base states} 
for the system? Now the question has been put incorrectly. There is no
such thing as ``{\em the}'' base states, because, of course, the
set of base states you may choose is not unique. New sets can always
be made out of linear combinations of the old. There are always many
choices for the base states, and among them, any choice is equally 
legitimate. So the question is not what is {\em the} base set, but
what {\em could} a base set be? We can choose any one we wish for our
own convenience. It is usually best to start with a base set which is
{\em physically} the clearest. It may not be the solution to any problem,
or may not have any {\em direct} importance, but it will generally
make it easier to understand what is going on.
\end{quotation}

The first set of basic states introduced there are
\begin{equation}
 \ket{\msl1} = \ket{{+}{+}},\quad
 \ket{\msl2} = \ket{{+}{-}},\quad
 \ket{\msl3} = \ket{{-}{+}},\quad
 \ket{\msl4} = \ket{{-}{-}}.
\label{set1}
\end{equation}

Feynman asked next
\begin{quotation}
\small
 You may say, ``But the particles interact, and maybe these aren't
right base states. It sounds as though you are considering the
two particles independently.'' Yes, indeed! The interaction raises
the problem, that is the {\em Hamiltonian} for the system, but 
interaction is not involved in the question, of how to {\em describe}
the system. What we choose for the base states has nothing to do
with what happens next. It may be that the atom cannot ever {\em stay}
in one of these base states, even if it is started that way.
That's another question. That's the question: How do the amplitudes
change with time in a particular (fixed) base? In choosing the base
states, we are just choosing the ``unit vectors'' for our description.
\end{quotation}

So, also other set of base states is introduced next \cite[\textbf{12-3}]{FLP3}
\begin{equation}
\begin{split}
 \ket{I} = \ket{{+}{+}},\quad
 &\ket{II} = \ket{{-}{-}},\quad
 \ket{III} = \frac{1}{\sqrt{2}}\bigl(\ket{{+}{-}}+\ket{{-}{+}}\bigr),\\
 &\ket{IV} = \frac{1}{\sqrt{2}}\bigl(\ket{{+}{-}}-\ket{{-}{+}}\bigr).
\end{split}
\label{setII}
\end{equation}
It is the set of stationary states, convenient for description
of the dynamics of the quantum system.

The citations above may be considered in relation with the
question: Why such active discussions about the entanglement was 
revived in quantum computing, if the ``regular'' quantum mechanics 
during about a half century managed almost without even mentioning 
of the term? 

Really, if choosing of the base states is matter of {\em convenience}, 
there is no big difference between $\ket{III}=(\ket{01}+\ket{10})/\sqrt{2}$ 
and $\ket{\msl3}=\ket{01}$ and so the term ``entanglement'' would
be rather redundant. It should be mentioned, that possibility of
using arbitrary basis for a {\em single} system seems quite clear, 
but here was emphasized an example with the basis ambiguity problem for 
the {\em compound system} and not with respect to only so-called 
{\em local transformations}, but for arbitrary change of basis in 
product space.

The theory of quantum algorithms may be considered as a modification 
of the theory of classical algorithms using specific artificial procedure of 
``quantization,'' then any discrete structure like the {\em bit} or the 
{\em array of bits} corresponds to the elements of basis in some Hilbert space. 
Such a construction {\em by definition} introduces some preferred bases, 
but it is just a problem discussed above, and it may contradict natural 
properties of quantum systems.

Of course, the notion of entanglement\footnote{In the theory of quantum 
communications most often used more difficult conception related with 
the {\em mixed states entanglement}, but it is not discussed in present note.} 
is very natural and convenient in the {\em mathematical} theory of quantum
computations, but the fact {\em alone} is not a guarantee of relevance to 
quantum mechanics.

Seems the ``na\"ive'' definition of compound system and permanent
discussion about entanglement as a mysterious and wonderful
phenomenon in quantum information science silently supposes 
rather classical point of view, that if we have two systems 
in the state $\ket{0}$, it is absolutely clear, that the state 
$\ket{0}\ket{0}$ is a ``natural'' and well defined state 
for a compound system, but the states like
$\ket{IV}=(\ket{01}-\ket{10})/\sqrt{2}$ 
suppose some weird manipulations with such classical states like $\ket{01}$
and $\ket{10}$.
 
It is not so, say for two spin-half systems the most natural state is just 
$\ket{IV}$, because it is so-called {\em singlet state} and most often 
it is the {\em ground state} of the quantum system. So it is strange 
to consider entanglement as demonstration of some nonusual properties
of the system --- it is much simpler to find or prepare hydrogen atom 
in the ground state $\ket{IV}$, than in any non-entangled one.

\end{document}